\documentclass{article}

\usepackage{arxiv}

\usepackage{amsmath,amssymb,amsfonts}
\usepackage{algorithmic}
\usepackage{graphicx}
\usepackage{textcomp}
\usepackage{makecell}
\usepackage[numbers]{natbib}
\usepackage{tabularx}
\usepackage{adjustbox}
\usepackage{soul}
\usepackage[normalem]{ulem}
\usepackage{multirow}
\usepackage{etoolbox}
\usepackage{pifont}
\usepackage{threeparttable}
\usepackage{xcolor}
\usepackage[linesnumbered,ruled,vlined]{algorithm2e}
\newcommand{\xmark}{\ding{53}}

\SetCommentSty{mycommfont}

\SetKwInput{KwInput}{Input}                
\SetKwInput{KwOutput}{Output}              

\def\tsc#1{\csdef{#1}{\textsc{\lowercase{#1}}\xspace}}
\tsc{WGM}
\tsc{QE}

\newtoggle{finalPaper}
\toggletrue{finalPaper} 
\iftoggle{finalPaper} {
	\newcommand{\addtxt}[1]{#1}
	\newcommand{\change}[2]{#2}
	\newcommand{\rmvtxt}[1]{}}
{
	\newcommand{\addtxt}[1]{\textcolor{red}{#1}}
	\newcommand{\change}[2]{\sout{#1}\textcolor{red}{#2}}
	\newcommand{\rmvtxt}[1]{\sout{#1}}}

\title{TemporalFED: A Software Module for Detecting Cyberattacks in Industry 4.0 Time-Series Data Using Decentralized Federated Learning}

\author{
\'Angel Luis Perales G\'omez\\
Department of Computer Technology and Engineering\\
University of Murcia\\
Espinardo, Murcia, Spain\\
\texttt{angelluis.perales@um.es}\\
\And
Enrique Tom\'as Martínez Beltr\'an\\
Department of Information and Communications Engineering\\
University of Murcia\\
Espinardo, Murcia, Spain\\
\texttt{enriquetomas@um.es@um.es}\\
\And
Pedro Miguel S\'anchez S\'anchez\\
Department of Information and Communications Engineering\\
University of Murcia\\
Espinardo, Murcia, Spain\\
\texttt{pedromiguel.sanchez@um.es}\\
\And
Alberto Huertas Celdr\'an\\
Communication Systems Group CSG\\
Department of Informatics IfI, University of Zurich\\
CH--8050 Zürich, Switzerland\\
\texttt{huertas@ifi.uzh.ch}\\
}

\begin{document}
\maketitle
\begin{abstract}
Industry 4.0 has brought numerous advantages, such as increasing productivity through automation. However, it also presents major cybersecurity issues, such as cyberattacks affecting industrial processes. Federated Learning (FL) combined with time-series analysis is a promising cyberattack detection mechanism proposed in the literature. However, having a single point of failure and network bottleneck are critical challenges that need to be tackled. Thus, this article explores the benefits of the Decentralized Federated Learning (DFL) in terms of cyberattack detection and resource consumption. The work presents TemporalFED, a software module for detecting cyberattacks in industrial environments using FL paradigms and time series. TemporalFED incorporates three components: time series conversion, feature engineering, and stationary conversion. To evaluate TemporalFED, it was deployed on Fedstellar, a DFL framework. Then, a pool of experiments measured the detection performance and resource consumption in a chemical gas industrial environment with different time-series configurations, FL paradigms, and topologies. The results showcase the superiority of the configuration utilizing DFL and Semi-Decentralized Federated Learning (SDFL) paradigms, along with a fully connected topology, which achieved the best performance in anomaly detection. Regarding resource consumption, the configuration without feature engineering employed less bandwidth, less usage of the Central Processing Unit (CPU), and less usage of the Random Access Memory (RAM) than other configurations.
\end{abstract}


\section{Introduction}

The fourth industrial revolution (Industry 4.0) encompasses the convergence of physical systems and digital technologies to create smart, interconnected, collaborative, and autonomous industrial processes \cite{ghobakhloo2020industry}. Key components of Industry 4.0 include the Internet of Things (IoT), artificial intelligence, cloud computing, big data analytics, and cyber-physical systems. These technologies allow industries to optimize operations, enhance productivity, reduce costs, and enable more efficient decision-making.

However, as industries increasingly adopt digital technologies and interconnected systems, they become more susceptible to anomalies provoked by cyberattacks, misconfigurations, or failures \cite{corallo2020cybersecurity}. Cybercriminals target industrial systems to disrupt operations, steal sensitive data, cause financial losses, and even compromise workers' safety. Common cyberattacks affecting industries include i) Ransomware, encrypting critical data or systems to demand a ransom; ii) Distributed Denial of Service (DDoS), disrupting industrial processes with a flood of traffic; iii) Insider Threats caused by employees performing malicious actions or stealing data; or iv) Supply Chain Attacks, targeting vulnerabilities in the supply chain to gain access to industrial systems.

Federated Learning (FL) has recently been proposed as a promising paradigm to detect most of the previous cyberattacks~\cite{Rey:2021:FL}. In the context of Industry 4.0, FL offers significant benefits in terms of collaboration and privacy preservation. More in detail, industrial organizations can collaborate on improving their cyberattack detection capabilities without sharing proprietary or sensitive data. Sensitive data remains on individual devices or industries while ML/DL model parameters are exchanged \cite{hao2019efficient}. However, a single point of failure and network bottleneck are critical drawbacks of centralized FL (CFL) in Industry 4.0 that must be avoided. In this sense, Decentralized Federated Learning (DFL)~\cite{MartinezBeltran:DFL_survey:2022} advances CFL by eliminating the need for a central server in the learning process.

Despite the advances of works proposing FL to detect cyberattacks in Industry 4.0, critical open challenges still need more effort \cite{tahir2021experience}. First, related work follows a CFL setup, where the central server is a single point of failure. Furthermore, the central server becomes a bottleneck during the aggregation process, especially as the number of devices increases. Second, as the literature identifies, analyzing time-series data is crucial for detecting anomalies caused by cyberattacks, creating unusual patterns or behaviors that deviate from the expected norms. However, there is no DFL-oriented work considering time series analysis to detect cyberattacks affecting Industry 4.0. Finally, there is no work evaluating the performance detection and resource consumption of DFL versus CFL setups using time series.  

In order to improve the previous challenges, the work at hand presents the following contributions:

\begin{itemize}

    \item The creation of a time-series analysis module, called TemporalFED, for Fedstellar, a well-known platform for training FL models in a decentralized fashion \cite{MartinezBeltran:fedstellar:2023}. TemporalFED consists of three components that are capable of processing time-series data. The first component \change{is responsible for converting}{converts} tabular data into time-series data. The second component \change{is in charge of extracting}{extracts} new features \change{by means of}{using} Discrete Fourier Transform (DFT) and autocorrelation (AC). Finally, the last component \change{is responsible for transforming}{transforms} non-stationary time-series data into a stationary one by removing trends and seasonality.
    
    \item The validation of TemporalFED in a simulated but realistic chemical gas industrial scenario. In particular, the testbed selected is the Tennessee Eastman (TE) process, which simulates chemical reactions in an industrial plant. This testbed is especially interesting in federated scenarios since it contains several simulations per normal and abnormal class. \rmvtxt{Furthermore, it enables the simulation of a geographically distributed industrial plant with similar industrial processes.}
    
    \item A pool of experiments evaluating the performance of the proposed component in the previous industrial scenario. The first\rmvtxt{ type of} experiment evaluated the anomaly detection performance of TemporalFED using several configurations and topologies of FL. In this experiment, the Semi-Decentralized Federated Learning (SDFL) paradigm with a fully connected topology achieved the best result with a 0.9406 F1-score, followed by the DFL paradigm using a fully connected topology with a 0.9405. In contrast, Centralized Federated Learning (CFL) achieved the worst results (varying the F1-score from 0.8762 to 0.9082). \rmvtxt{The second type of experiment measured the resource consumption of the TemporalFED in terms of CPU and RAM usage and bytes exchanged by the federation participants. In this experiment, the TemporalFED configuration that did not use the feature engineering component consumes more resources than the others. It is worth mentioning that the CFL paradigm exchanges more bytes between participants.}

    \item \addtxt{An experiment in charge of measuring the resource consumption of the TemporalFED in terms of CPU and RAM usage and bytes exchanged by the federation participants. In this experiment, the TemporalFED configuration that did not use the feature engineering component consumes more resources than the others. It is worth mentioning that the CFL paradigm exchanges more bytes between participants.}
    
\end{itemize}

The remainder of this article is structured as follows. Section \ref{sec:related_work} gives an overview of FL-based solutions detecting cyberattacks in industrial scenarios. Section \ref{sec:module} describes Fedstellar and the time series analysis component to detect cyberattacks in industrial environments. Section \ref{sec:experiments} presents the experimental results when detecting cyberattacks in different federated scenarios. Finally, Section \ref{sec:conclusions} gives an overview of the conclusions extracted from the present work and future research directions.

\section{Related Work}
\label{sec:related_work}

This section analyzes which solutions are available for decentralized and time series-based model generation. In this sense, it seeks to identify the need for a platform capable of integrated DFL and time series processing. Afterward, this section reviews the state-of-the-art works applying FL for time series in industrial cyberattacks and fault detection.

\subsection{DFL solutions in industrial scenarios}

With Industry 4.0 maturing into 5.0, the concept of DFL is becoming a pivotal element in handling the vast volumes of data produced and processed within industry domains. Martínez et al. \cite{MartinezBeltran:DFL_survey:2022} comprehensively reviews the DFL landscape, addressing its fundamentals, applications, existing solutions, and associated challenges. In this work, the authors thoroughly explore DFL as a solution, underling its advantages in minimizing data sharing and avoiding centralized architecture dependencies. Their findings are instrumental in identifying the limitations of existing DFL solutions for industrial scenarios, forming the basis for creating novel functionalities suited for these contexts.

Among the numerous practical applications of DFL, one area of interest is the implementation of decentralized FL systems in industrial wireless networks. Savazzi et al. \cite{Savazzi:dfl_industry_networks:2020} provide a real-time framework for analyzing DFL running on top of industrial networks rooted in the IEEE 802.15.4e standard. This research contributes to optimizing network and ML model deployment, with special attention to model pruning, sparsification, and quantization. While DFL effectively addresses a single point of failure, the limitations of communications resources in Device-to-Device (D2D) links present another challenge. Ma et al. \cite{Ma:iot_selection_quantized:2021} introduce a quantization-based DFL (Q-DFL) mechanism, demonstrating its effectiveness in reducing data transmission volume and ensuring the convergence of DFL models. Moreover, they designed a consensus mechanism and synchronizing protocols to manage transitions between phases in the industrial field.

In the realm of Industry 4.0, cognitive computing introduces another layer of complexity to the automation process. Qu et al. \cite{Qu:bl_cognitive_computing_industry:2021} propose a decentralized paradigm merging FL and blockchain to resolve performance and security issues. This approach leverages the strength of blockchain-enabled FL, promoting quick convergence with advanced verifications and member selections. The advent of Industry 5.0 intensifies the issues related to centralization, privacy preservation, latency, and security in industrial infrastructures. Singh et al. \cite{Singh:bl_5g_dfl:2023} present FusionFedBlock, a fusion of blockchain and FL framework designed to uphold privacy in Industry 5.0. It demonstrates consistent performance using CIFAR-10 and FEMNIST datasets, achieving a notable 93.5\% accuracy in active nodes. Renathunga et al. \cite{Ranathunga:dfl_bl_industry:2023} present a similar solution using a hierarchical network of aggregators to evaluate the quality of local model updates, offering a flexible and efficient solution that outperforms current state-of-the-art methods in terms of accuracy, throughput, and latency.

Continuing this trend in industrial scenarios, Yuan et al. \cite{Yuan:decefl_framework:2021} propose a DeceFL framework, which operates independently of a central client, relying solely on local transmissions. The proposed framework showcases its effectiveness in both convex and nonconvex loss functions, time-invariant and time-varying topologies, and IID and Non-IID datasets. The authors validated the solution with a time-series dataset. However, the approach applied did not treat the data as truly time series since required processing of such data type is missing. Similarly, Qui et al. \cite{Qiu:dfl_industry:2023} focus on the application of DFL in the context of the Industrial Internet of Things (IIoT). In an effort to counter the threat of data leakage and misuse by central servers, the authors propose a DFL algorithm based on deep neural networks. The algorithm employs an intrinsic plasticity method that addresses the problem of unreliable central servers in the peer-to-peer structure. In addition, it combines the decentralized medium consensus and the alternating direction algorithm of multipliers, which each agent solves with its own local data.

\subsection{FL and time series for industrial cyberattack detection}

From a wider perspective and not only focusing on decentralized setups, Rodríguez et al. \cite{rodriguez2023anomaly}, and Nain et al. \cite{nain2022towards} revised the most recent literature on anomaly detection and edge computing in industrial scenarios. Besides, Hiessl et al. \cite{hiessl2020industrial} revised the requirements and system designs for industrial FL applicability. These works highlight the importance of time series processing and FL in the state-of-the-art. However, they also remark on challenges related to the lack of processing tools for direct scenario evaluation. 

Liu et al. \cite{liu2020deep} were among the first authors that proposed the usage of FL for anomaly detection in industrial time-series data. They used a convolutional Long-Short Term Memory (LSTM) network that leverages attention mechanisms. Gradient compression was also applied to reduce 50\% of the communication overhead introduced by the centralized FL aggregation mechanism. Four real-world datasets were employed for validation, demonstrating the good performance of the approach.

Huong et al. \cite{huong2021detecting} deployed a combination of Variational Autoencoder (VAE) and LSTM to cope with anomaly detection. Several datasets were employed for validation, improving literature results in most of them. Besides, they verified that the resource usage of the approach is suitable for Industrial IoT devices. Similarly, Truong et al. \cite{truong2022light} also applied FL for anomaly detection for time-series data in ICS. In this case, the authors centered their solution on the lightweight perspective. To that end, they combine Fourier transform operation as preprocessing with Transformer Autoencoder models for anomaly detection. The advantage of these models against recurrent ones, such as LSTM, is their faster training due to the absence of recursion during training. This approach was also evaluated with several datasets of industrial scenarios, such as electricity, gas, and water supply control. Huu Du et al. \cite{du2023trans} also proposed FL and transformers applied to ICS, in this case, for remaining lifetime prediction. They were able to improve state-of-the-art by up to 25\% using this attention-based mechanism.

Jahromi et al. \cite{jahromi2023ensemble} implemented an ensemble-based solution for federated threat detection in IIoT. Two different models are combined, one looking for the normal functioning of the devices and another checking about threat identification. For validation, the SWaT and gas pipeline datasets were selected, achieving better performance than other solutions in the literature. Also, for intrusion detection, De Carvalho Bertoli et al. \cite{de2023generalizing} proposed a generalist unsupervised FL approach using stacked ensemble models. This idea was validated using several datasets, some of them containing IIoT network traces. 

From other innovative perspectives, not directly leveraging Fl, Bin Masood et al. \cite{masood2023blockchain} employed Blockchain technologies for decentralized information sharing of faults in ICS. A DL-based model is employed as an observer of the faults in order to identify the component generating them. The approach is validated using the Tennessee Eastman Process (TEP) dataset. Moreover, Radaideh et al. \cite{radaideh2022time} also applied time series and LSTM networks for anomaly detection in high-voltage converters. Convolutional LSTM networks achieved up to 91\% precision, 88\% recall, and 20\% False Negative Rate. However, this work did not exploit a FL-based setup. From a different perspective, Zhang et al. \cite{zhang2022r} proposed the usage of reinforcement learning in the model fusion process, increasing accuracy and FL attack resilience in an industrial scenario.

\tablename~\ref{tab:related} compares the solutions revised in this section. From the list of works reviewed, some conclusions and open challenges can be extracted. First, it can be seen that industrial scenarios have not been extensively covered using DFL-based setups. Some works have introduced DFL in the industry, but still use rather simple datasets not based on real-world deployments. In contrast, FL and time series have been extensively applied in real-world industrial scenarios from a wide variety of perspectives, being the usage of attention-based mechanisms the trend nowadays. From these two takeaways, it arises the need for a platform capable of applying time-series analysis over DFL in industrial scenarios, making it possible to extrapolate traditional FL approaches to fully decentralized ones, which are the trend nowadays due to their advantages.

\begin{table}[ht!]
\centering
\caption{DFL, FL, and time series approaches for industrial scenarios}
\begin{tabular}{lccccl}
\hline
\textbf{Solution} & \makecell[c]{\textbf{Realistic}\\\textbf{Dataset}}  & \textbf{DFL} &  \textbf{FL} & \makecell[c]{\textbf{Time}\\ \textbf{Series}} & \makecell[c]{\textbf{Model or}\\\textbf{Framework}}\\
\hline
\cite{Savazzi:dfl_industry_networks:2020} 2020 & \checkmark & \checkmark & \xmark & \xmark & Framework\\
\hline
\cite{liu2020deep} 2020 & \checkmark & \xmark & \checkmark & \checkmark & \makecell[c]{Attention\\ LSTM} \\
\hline
\cite{Ma:iot_selection_quantized:2021} 2021 & \xmark & \checkmark & \checkmark & \xmark & Framework \\
\hline
\cite{Qu:bl_cognitive_computing_industry:2021} 2021 & \checkmark & \checkmark & \checkmark & \xmark & Blockchain\\
\hline
\cite{Yuan:decefl_framework:2021} 2021 & \checkmark & \checkmark & \checkmark & \xmark & Framework\\
\hline
\cite{huong2021detecting} 2021 & \checkmark & \xmark & \checkmark & \checkmark & VAE LSTM \\
\hline
\cite{radaideh2022time} 2022 & \checkmark & \xmark & \xmark & \checkmark & Conv LSTM \\
\hline
\cite{truong2022light} 2022 & \checkmark & \xmark & \checkmark & \checkmark & Attention \\
\hline
\cite{zhang2022r} 2022 & \checkmark & \xmark & \checkmark & \xmark & \makecell[c]{Reinforcement\\Learning}\\
\hline
\cite{Singh:bl_5g_dfl:2023} 2023 & \xmark & \checkmark & \checkmark & \xmark & Blockchain \\
\hline
\cite{Ranathunga:dfl_bl_industry:2023} 2023 & \xmark & \checkmark & \checkmark & \xmark & Blockchain\\
\hline
\cite{Qiu:dfl_industry:2023} 2023 & \xmark & \checkmark & \checkmark & \xmark & Algorithm\\
\hline
\cite{du2023trans} 2023 & \checkmark & \xmark & \checkmark & \checkmark & Attention \\
\hline
\cite{jahromi2023ensemble} 2023 & \checkmark & \xmark & \checkmark & \checkmark & Ensemble  \\
\hline
\cite{de2023generalizing} 2023 & \checkmark & \xmark & \checkmark & \checkmark & Ensemble \\
\hline
\cite{masood2023blockchain} 2023 & \checkmark & \xmark & \xmark  & \xmark & Blockchain \\
\hline
\textit{This work} & \checkmark &\checkmark & \checkmark & \checkmark & \makecell[c]{Framework,\\ LSTM} \\
\hline
\end{tabular}%
\label{tab:related}
\end{table}

\section{TemporalFED Design}
\label{sec:module}

This section details the design of the TemporalFED module that allows Fedstellar to process time-series data and detect cyberattacks in industrial scenarios. To fully understand this section, it starts with a brief overview of the Fedstellar platform, which is a general platform for DFL.

\subsection{Fedstellar}

Fedstellar is a platform that streamlines the training of Decentralized Federated Learning (DFL) models across various devices, industries, and organizations \cite{MartinezBeltran:fedstellar:2023}. \addtxt{It supports complex network configurations for DFL, SDFL, and CFL, enabling diverse network topologies. Additionally, the platform facilitates the deployment of physical or virtual participants and accommodates various ML/DL models and datasets for addressing FL challenges.}

\addtxt{Fedstellar offers user-friendly tools to adjust settings and monitor key metrics, including resource metrics (e.g., computing power, memory usage, data transfer speed) and performance metrics (e.g., accuracy, precision, recall, $F_1$ score, training time, and model improvement speed).}

\rmvtxt{The Fedstellar architecture employs an elaborate design that includes advanced data manipulation, effective training methodologies, and communication protocols. It comprises multiple functional elements to enable the deployment of different FL architectures across physical and virtual participants:}

\addtxt{The architecture of the platform integrates advanced data handling, training methods, and communication protocols. It consists of the following components:}

\begin{itemize}
\item \textit{Frontend.} \rmvtxt{This component is the interface where the user defines and watches over learning scenarios. It is designed to be easy to use, making system setup simple and tracking metrics easy.}\addtxt{A user-friendly interface for defining and tracking learning scenarios.}
\item \textit{Controller.} \rmvtxt{It is the control center of the platform. It takes user commands from the front end, manages the whole federated scenario, selects learning algorithms and datasets, and deploys network connections to ensure a smooth FL process.}{The orchestrator that manages the FL, selects algorithms and datasets, and configures network connections.}
\item \textit{Core.} This component is deployed on each participant in the federation. \rmvtxt{It is responsible for doing the FL tasks. It handles training models, preparing data, securing participant communication, and storing the federated models. It also keeps track of the important metrics and sends this information back to the front end for monitoring.}{It manages training, data preparation, secure communication, and storage of federated models while reporting metrics to the frontend.}
\end{itemize}

\subsection{TemporalFED: Time-Series Module}

Most industrial systems generate data with temporal dependencies, meaning that the values at time $t$ depend on the values at time $t-1$. It is crucial to consider this property to develop anomaly detectors that efficiently detect cyberattacks in industrial settings. However, in its current state, Fedstellar only accepts tabular data as input, thus necessitating the development of a specific module capable of handling and processing data in the form of time series.

The proposed module in this work consists of three interconnected components that work together to prepare the data for proper interpretation by the inference models implemented in Fedstellar. Specifically, the first component is designed to handle tabular data and convert it into a time-series format. The second component extracts new features that aid the anomaly detector in distinguishing between normal system behavior and behavior under attack. Finally, the third component is responsible for studying the trend and seasonality in the data and, if present, eliminating them to convert the time series into stationery. \addtxt{The separation between the Feature Engineering component and the Stationary Conversion component is due to better organization of the TemporalFED module. Although in some cases stationary conversion may be considered part of the feature engineering process, we have chosen to separate the two components so that feature engineering focuses on creating new features, while stationary conversion is responsible for modifying the existing features when necessary. The three components can be used either together or independently. For this, it is necessary to adapt the fedstellar front-end to allow such selection. This is done in the Time Series Definition component of the front-end. Essentially, this component provides the previously discussed techniques and a selector to choose one or several techniques to apply. Once the techniques to be used are selected, the Time Series Techniques component of the Controller will be responsible for applying the different techniques by utilizing the TemporalFED module.}

\begin{figure*}[htb]
	\centerline{\includegraphics[width=0.75\textwidth]{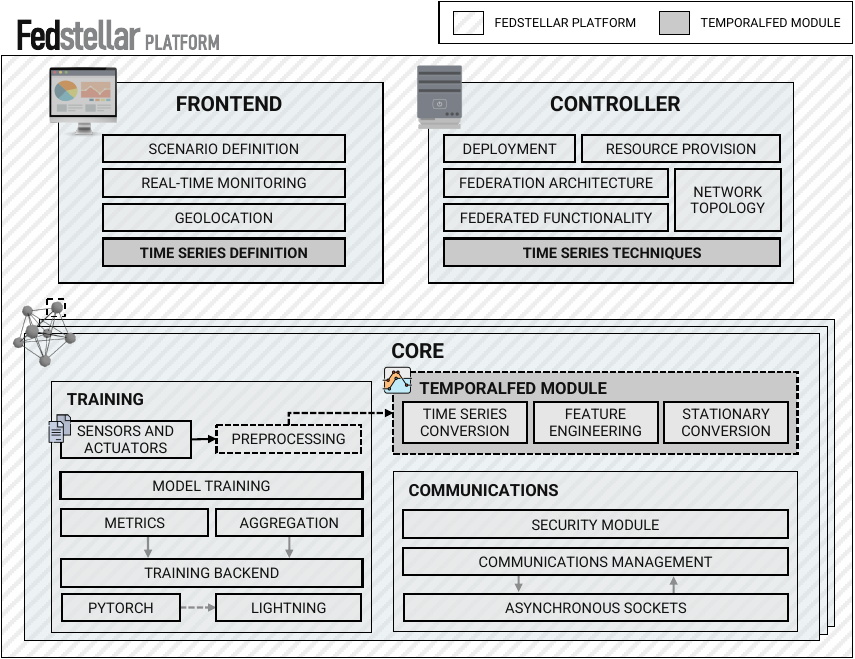}}
	\caption{High-level architecture of the time-series module}
	\label{fig:ts_module}
\end{figure*}

Figure~\ref{fig:ts_module} illustrates a high-level overview of the module and its components, as well as the integration into Fedstellar. The following sections will provide detailed explanations of the module components.

\subsubsection{Time series conversion}

This component is responsible for receiving input data in tabular format and converting it into a time-series format. Formally, the component takes data in the form $(m, n)$, where $m$ represents the available samples, and $n$ represents the features for each sample, and transforms it into a new dataset with the shape $(m, ts, n)$. Here, $m$ and $n$ retain the same meaning as in the previous case, while $ts$ denotes the selected time step, indicating the number of samples considered within each time unit. \addtxt{This component waits to receive the data from each participant in the federation. Once the data is received, it is transformed for further processing in the form of a time series. After the data has been transformed, if necessary, it will move on to the next component of TemporalFED.}

\subsubsection{Feature Engineering}

Typically, industrial devices are involved in controlling and monitoring equipment, often carrying out repetitive actions. As a result, this component employs two techniques to extract relevant characteristics that leverage this behavior. Initially, it utilizes the autocorrelation (AC) function on a specific window to extract features of higher orders. The formal expression of the AC is presented in Equation~\ref{eq:ac}.

\begin{equation}
    \label{eq:ac}
    autocorr_{x_w,k} = \frac{\sum_{i=w-W+1}^{w-k}(x_i-\hat{x})(x_{i+k}-\hat{x})}{\sum_{i=w-W+1}^{w}(x_i-\hat{x})^2}
\end{equation}

Where $k$ is the lag introduced, $W$ is the length of the window, $x_w$ is the value of $x$ at instant $w$, and $\hat{x}$ is the mean of the $x$ values in the window.

Subsequently, the component employs the Discrete Fourier Transform (DFT) on a designated window to extract features of higher orders. Specifically, the DFT offers a convenient method for converting a signal from the time domain to the frequency domain. The formal representation of the DFT can be observed in Equation~\ref{eq:dft}.

\begin{equation}
    \label{eq:dft}
    \overline{DFT}_{x_w,k} = \sum_{j=w-W+1}^{w}x_je^{-\frac{2\pi i}{W}k(j-(w-W+1))}
\end{equation}

where $x_w$ is the value of $x$ at instant $w$ and $W$ is the total number of samples.

After applying the techniques above to a window, the proposed approach considers the dominant values obtained from each technique. Specifically, the dominant values derived from the AC technique provide insights into the similarity between the current time-series signal and its future instances. On the other hand, the dominant values or frequencies extracted through the DFT technique represent the frequencies with the highest amplitudes in the signal. In the context of industrial scenarios, it is reasonable to assume that anomalies resulting from cyberattacks would alter these dominant values.

\addtxt{This component, like the previous one, is executed in each of the federation's participants. Specifically, it takes the data processed by the previous component and generates new features. Once these features are generated, they, along with the original features, are passed on to the next component of TemporalFED. Note that this component generates new features for each existing feature. Therefore, in a Federated Learning paradigm like the one used in this work, this implies an increase in data transmission when using this component, regardless of the federated learning topology employed.}

\subsubsection{Stationary conversion}
\label{sec:stationary_conversion}

Removing trends and seasonality from time-series data is crucial for several reasons. On the one hand, trends refer to long-term changes in the data, which can hide the underlying patterns and make it difficult to discern the true relationships within the data. By removing trends, the resultant model can focus on the inherent fluctuations and patterns of interest for analysis, allowing for a more accurate understanding of the underlying dynamics.

On the other hand, seasonality refers to repetitive patterns that occur within specific time intervals, such as daily, weekly, or yearly cycles. These patterns can introduce significant variability and distort the statistical properties of the data. Eliminating seasonality can reduce the impact of these repetitive patterns and obtain a clearer picture of the underlying dynamics that are not influenced by specific time intervals. This is particularly important for making reliable forecasts, as failing to account for seasonality can lead to inaccurate predictions and misleading conclusions.

Achieving stationarity by eliminating trends and seasonality in a time series is vital for applying various statistical models and analysis techniques. It ensures that the statistical properties of the series, such as mean, variance, and autocorrelation, remain constant over time, allowing for more robust and accurate analysis. By obtaining a stationary time series, researchers can make informed decisions, develop reliable models, and draw valid conclusions based on the data, leading to advancements in scientific understanding and practical applications in various fields.

This work proposes to use simple and efficient techniques to detrend and convert time series generated in industrial environments into stationary ones. To do this, this component first checks if the input time-series data is stationary. For this, the component uses the Augmented Dickey-Fuller (ADF), a statistical test used to determine whether a time series is stationary or non-stationary. It compares the coefficient of the lagged differenced variable in an autoregressive model against critical values. If the test statistic is less than the critical value, the null hypothesis of non-stationarity is rejected, indicating stationarity. On the other hand, if the test statistic is greater than the critical value, the null hypothesis is not rejected, suggesting non-stationarity. The ADF test accounts for serial correlation and higher-order autoregressive processes, making it a reliable method to assess stationarity in time-series data.

If the ADF test returns that the time series is stationary, this component does no additional work. Otherwise, the component proceeds to eliminate the trend and seasonality. First, the component uses a simple but runtime-efficient technique to remove the trend. In particular, it is proposed to use the difference between the time series and the moving average of the time series. With this simple operation, the trend is eliminated from the time series. Second, to eliminate the seasonality of the time series, the component uses the differentiation technique, which is another simple and efficient technique in terms of execution. This technique calculates the difference between the current value and its value in the previous season. However, to apply this technique correctly, it is necessary to know the repetition period of the season. Simple patterns can be identified by plotting the data. However, this is not scalable, automatic, or works in cases where the patterns are more complex. Therefore, the component uses AC to determine such parameters. Once the component knows the frequency with which the season is repeated, the differentiation technique can be applied to eliminate the seasonality. However, because the problem faced is anomaly detection, this component only examines and eliminates the season and trend in the normal data. The reason is that a cyberattack can produce such patterns, and if the component eliminates them from abnormal data, it loses a huge amount of information.

\addtxt{Within the federation, this component is executed by each participant. Specifically, it receives the processed data from the previous component and checks if the original dataset is stationary. Once the data has been converted to stationary, it is passed to the model training module of Fedstellar, which is executed within each participant in the federation.}

\subsubsection{\addtxt{Interaction between components}}

\begin{figure*}[tp]
	\centerline{\includegraphics[width=0.6\textwidth]{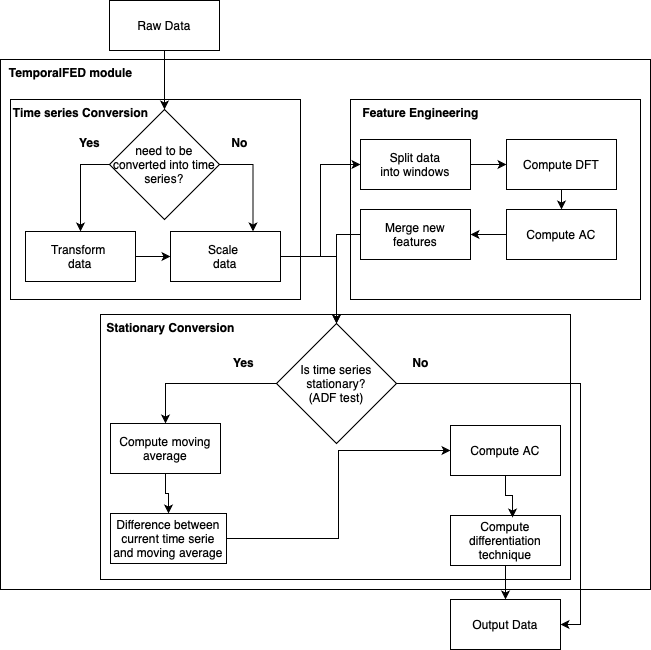}}
	\caption{Diagram showing how each TemporalFED component works}
	\label{fig:diagram}
\end{figure*}

\addtxt{This section briefly outlines the interactions between the input data and each component of TemporalFED. The graphical representation of these interactions can be seen in~\figurename~\ref{fig:diagram}.}

\addtxt{First, the raw data held by the participant is passed to the Time Series Conversion component. If the data needs to be converted into time series format, it is transformed and scaled. Otherwise, the data is simply scaled and then provided to the subsequent components of TemporalFED. Once the data has been transformed into a time series, it is processed simultaneously by the Feature Engineering and Stationary Conversion components, as there are no dependencies between them and both work with the original features.}

\addtxt{On one hand, when the data reaches the Feature Engineering component, it is divided into windows, the DFT and AC are computed, and the new features are combined with the original ones. On the other hand, in the Stationary Conversion component, the original data is first checked to see if it is stationary. If so, a moving average of the feartures is computed and subtracted from the original ones to remove the trend. Next, the AC is calculated to detect seasonality, and differentiation is applied to eliminate it. This modifies the non-stationary samples, and the results, along with the output from the Feature Engineering component, form the final output data.}

\section{Experiments}
\label{sec:experiments}

This section presents the experiments to validate the novel TemporalFED module integrated into Fedstellar. First, it details the industrial testbed where the dataset used during the experiments was collected. Then, it depicts the configuration needed in both Fedstellar and the time-series component. In particular, the DL model employed is described together with its hyper-parameters, as well as the participants in the federation, the topology, the data preprocessing, and the specific configuration of the AC, DFT, and the techniques to remove non-stationary data. Finally, this section shows the results in multi-class classification obtained using several of the configurations explained before. Furthermore, it also presents experimentation in terms of resource consumption testing different paradigms, topologies, and time-series module configurations.

\subsection{Testbed}
\begin{figure*}[tp]
	\centerline{\includegraphics[width=0.75\textwidth]{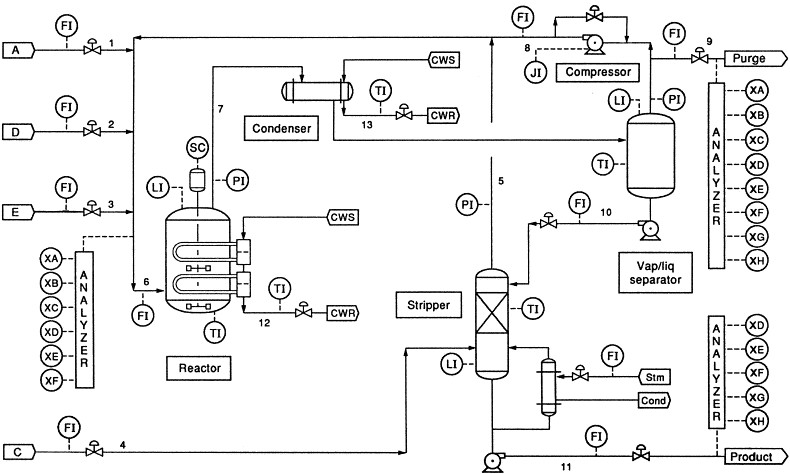}}
	\caption{Tennessee Eastman Process}
	\label{fig:tep}
\end{figure*}

The TE process is illustrated in Fig.~\ref{fig:tep}, showcasing five primary modules: the reactor, condenser, liquid-vapor separator, compressor, and stripper. This process generates two products by reacting eight components: A, B, C, D, E, F, G, and H. The following equations define the reaction performed in the testbed:

\begin{align*}
A(g) + C(g) + D(g) \rightarrow G(liq) \qquad Product 1\\
A(g) + C(g) + E(g) \rightarrow H(liq) \qquad Product 2 \\
A(g) + E(g) \rightarrow F(liq) \qquad Byproduct \\
3D(g) \rightarrow 2F(liq) \qquad Byproduct
\end{align*}

During the process, input reactants A, D, E, and C are introduced into the reactor, where some of the aforementioned reactions occur. This leads to vapor products and unreacted components forming, which flow into the condenser. These products and unreacted components undergo cooling and transition from gaseous to liquid states in the condenser. The resulting mixture is then directed to a separator that separates the unreacted components, which are subsequently recycled and reintroduced into the process through the compressor. On the contrary, the condensed products are directed to a product stripping module, where any remaining reactants are removed. Finally, the output of the stripper yields the production of products G and H.

In addition, all the reactions occurring in the process are irreversible and release heat, and they exhibit a nearly first-order dependence on the concentrations of the reactants. The rates at which these reactions proceed depend on the temperature and follow an Arrhenius expression. Among the various components, G is particularly sensitive to temperature changes due to its higher activation energy.

The control objectives for this process are the following:

\begin{itemize}
    \item Control process variables to keep them at desired values.
    \item Ensure that the operational parameters of the process remain within the limitations of the equipment.
    \item Reduce the fluctuations in both the rate and quality of the product when unexpected disruptions occur.
    \item Reduce the need for frequent valve adjustments that may impact other processes.
    \item Efficiently and promptly recover from disturbances, changes in production rate, or adjustments in the product mix while maintaining smooth operations.
\end{itemize}

The dataset used during the following sections was generated by Rieth et al.~\cite{rieth2017issues} using the testbed TE process detailed in this section. The authors released four files: training files and test files containing anomalies, as well as training files and test files without any anomalies. Each file consists of 500 simulations for each type of anomaly. In the training files, each simulation contains 500 samples, while the test files comprise 960 samples. These files encompass 52 features, consisting of 41 measurement variables and 11 manipulated variables sampled every 3 minutes. This results in 25 hours for the training datasets and 48 hours for the test datasets.

One of the reasons for using this dataset is that it provides multiple simulations for each anomaly type and normal class. This allows us to simulate different industrial plants located in different geographical areas, all carrying out a similar industrial process. As a result, it enables us to conduct experiments in a realistic environment, not only at an industrial level but also in terms of FL.

\subsection{Fedstellar and time-series component configuration}

This section provides a detailed explanation of all the configurations required to conduct the experiments.

Firstly, the dataset extracted from TEP was divided into training, validation, and test sets. The training set included the initial 60 simulations, comprising a total of 480 samples from the training files. The first 20 samples were excluded since they were mislabeled as normal in the anomaly simulations. The validation set consisted of the subsequent 30 simulations from the training files, each containing 480 samples. Similarly, the first 20 samples were removed for each simulation. Finally, the test set comprised the first 30 simulations from the test files. The initial 160 samples were discarded as they were labeled as normal for the anomaly simulations. Therefore, for each simulation, samples ranging from 161 to 500 were selected, resulting in a total of 340 samples. Furthermore, the datasets were normalized using a standard scaler that calculates the mean and standard deviation of the training set and utilizes them to normalize the evaluation and test sets. Subsequently, the samples referring to anomalies 3, 9, and 15 were eliminated since they did not suffer enough variation to be considered anomalies, as pointed out by~\cite{onel2019nonlinear}. 

The time-series conversion component was fixedly configured to have a temporal sequence of 5 elements. Thus, the dataset was shaped as $(m, 5, n)$. The values of $m$ and $n$ varied depending on the dataset (training/validation/test) and whether the DFT and AC components were employed. Specifically, each federation participant utilizes 10 simulations for training and 5 simulations for evaluation and testing. This resulted in a total of 68\,400 samples for training, 43\,200 for evaluation, and 30\,600 for testing. As for the value of $n$, it depended on whether the DFT/AC component is used. If this component was not employed, $n$ took the value of 52, which corresponds to the provided dataset features. On the other hand, if the component was activated, 4 additional features are extracted for each original feature (2 using DFT and 2 using AC), resulting in $n$ taking the value of 260. Lastly, in some of the experiments, the component that transforms non-stationary series into stationary ones using the technique described in Section~\ref{sec:stationary_conversion} was activated.

Regarding the DL model used in Fedstellar during the experiments, it was an LSTM network. Specifically, it was a two-layer LSTM network consisting of an input layer, two hidden layers, and an output layer. The first hidden layer has 128 neurons, while the second has 64 neurons. The output layer was a dense network with 21 neurons. The chosen activation function for the hidden layers was ReLU, and for the output layer, Softmax was employed. As a result, the output provides the probability for each of the considered classes. Finally, the batch size was set to 1024, and both rounds and epochs were fixed to 10. Table~\ref{tab:lstm_architecture} summarizes the architecture of the DL model. This work did not consider carrying out a grid or random search to find the optimal hyper-parameters since this work aims to probe the feasibility of Fedstellar to detect anomalies in an industrial setting, not to find the best DL model. 

\begin{table}[]
\centering
\caption{Architecture of LSTM model used to detect anomaly detection}
\label{tab:lstm_architecture}
\begin{tabular}{ll}
\hline
\multicolumn{1}{c}{\textbf{Layers}} & \multicolumn{1}{c}{\textbf{Hyper-parameters}} \\ \hline
LSTM Layer                            & Neurons: 128                                   \\ \hline
Activation                            & ReLU                                   \\ \hline
LSTM Layer                            & Neurons: 64                                   \\ \hline
Activation                            & ReLU                                   \\ \hline
Dense Layer                           & Neurons: 21                                   \\ \hline
Activation                               & Softmax                                      \\ \hline
Loss                           & Categorical crossentropy                                    \\ \hline
Epochs                               & 10                                      \\ \hline
Batch size                           & 1024                                    \\ \hline
Optimizer                           & Adam                                    \\ \hline
\end{tabular}
\end{table}

Finally, to conduct the experiments, Fedstellar is configured using different topologies and paradigms of FL. Specifically, it was considered the fully connected, ring, and star topologies. Regarding the paradigms of FL, it was considered DFL, SDFL, and CFL. In this context, DFL and SDFL were tested with the fully connected and ring topologies, while CFL, by design, used star topology. The method employed by Fedstellar to aggregate the weights of the model was FedAvg, a popular technique that consists in performing the average of the weights of all participants in the federation \cite{McMahan:communication_efficient:2016}.

With the aforementioned configurations, various combinations were conducted to determine the optimal setup, not only in terms of anomaly detection performance but also considering the resource consumption (\% CPU used, \% RAM used, and number of bytes transmitted). During these experiments, the TemporalFED was configured in different ways. The first configuration enabled only the component that transforms the tabular data into time series. This component is also enabled in the subsequent configuration since it is mandatory to process time-series data. The second configuration activated all three components of the TemporalFED. The third configuration activated the feature engineering component. Finally, the fourth configuration enabled the stationary conversion component. Considering these four configurations, the three different topologies (fully connected, star, and ring), and the three considered paradigms (DFL, SDFL, and CFL), a total of 20 experiments have been performed.

The experiments were deployed in a workstation with 128GB of RAM, a six-core Intel i9-12900K at 3.2GHz with hyper-threading running Linux. As mentioned previously, we deployed 5 participants with identical CPU/RAM resources. Besides, the Fedstellar models were deployed in the CPU.

\subsection{Metrics}

This section provides a comprehensive explanation of the metrics utilized to evaluate the performance of the anomaly detection model, both in terms of anomaly detection and resource consumption.

To assess the performance of the anomaly detection model, it is common to employ metrics such as precision, recall, and F1-score, which are calculated based on the concepts of True Positive (TP), True Negative (TN), False Positive (FP), and False Negative (FN):

\begin{itemize}
\item True Positive (TP): Represents the number of correctly detected anomalies by the model.
\item True Negative (TN): Represents the number of non-anomalies correctly classified.
\item False Positive (FP): ndicates the number of non-anomalies mistakenly classified as anomalies.
\item False Negative (FN): Indicates the number of anomalies mistakenly classified as non-anomalies.
\end{itemize}

Subsequently, precision, recall, and F1-score are defined as follows:

\begin{itemize}
\item Reflects the proportion of correctly detected anomalies among all the detected anomalies. It is calculated as follows: $$ \mbox{\textit{Precision}} = \frac{TP}{TP +FP}$$
\item Recall: Indicates the proportion of actual anomalies that are correctly identified by the model. It is calculated as follows: $$ \mbox{\textit{Recall}} = \frac{TP}{TP +FN}$$
\item F1-score: Represents the harmonic mean of precision and recall, offering a balanced trade-off between the two. It is calculated as follows: $$ \mbox{\textit{F1-score}} = 2 \times \frac{Precision \times Recall}{Precision +  Recall}$$
\end{itemize}

In terms of resource consumption, the number of bytes transmitted was measured, which is computed as the sum of the bytes received and sent, the CPU percentage used, and the percentage of RAM required to perform each experiment. It is worth mentioning that each experiment involved 5 participants. However, for simplicity in presenting the data below, the results for each experiment represent the average of the 5 participants' outcomes.

\subsection{Results}

In this section, the results obtained from executing the experiments discussed in the previous section are presented. Firstly, the results pertaining to each experiment's anomaly detection performance are presented. Secondly, the resource consumption produced by each launched experiment will be detailed.

\tablename~\ref{tab:ad_results} displays the results concerning anomaly detection performance. An interesting result is that the CFL paradigm yields the poorest results for any configuration of the TemporalFED module. Specifically, the maximum F1-score achieved within these configurations is 0.9082, while the minimum is 0.8762. Despite exhibiting the worst performance, CFL can still be considered suitable for industrial environments where the patterns in transported data can be exceedingly complex. However, this reinforces the utility of employing a distributed approach (DFL and SDFL) for anomaly detection in industrial settings.

Regarding optimal precision performance, the DFL paradigm with a fully connected topology and solely activating the time-series conversion component achieved the highest score with 0.9509. Regarding the recall, the approach utilizing the SDFL paradigm, a fully connected topology, and only the time-series conversion component attained the best performance (0.9370). Furthermore, this same approach also achieved the highest F1-score (0.9406), thus exhibiting superior overall performance in anomaly detection.

Considering the other configurations of the TemporalFED module, it is evident that the use of feature engineering component based on DFT and AC slightly degrades performance for all tested topologies and paradigms. Additionally, utilizing all components of the TemporalFED module also resulted in slightly lower performance due to the activation of the feature engineering component using DFT and AC.

In connection with the configuration using the stationary conversion component, promising results were observed, outperforming other paradigms and topologies in certain cases. For instance, when employing DFL with a ring topology, this approach achieved a precision score of 0.9460, surpassing the configuration using only the conversion to time series (0.9414). This same pattern was also observed for recall and F1-score metrics for the aforementioned configurations, with the DFL, ring, and stationary conversion component attaining a recall of 0.9324 and an F1-score of 0.9364. In contrast, the DFL, ring, and time-series conversion achieved a recall of 0.9266 and an F1-score of 0.9306.

In conclusion, the results indicate that the distributed approaches, particularly utilizing the DFL and SDFL paradigms, improve performance in detecting anomalies in complex industrial environments. Moreover, careful consideration of the configuration of the TemporalFED module and the selection of relevant components significantly impact the overall detection performance.

\begin{table*}[]
\centering
\caption{Performance in anomaly detection achieved by each TemporalFED configuration and under different paradigms and topologies}
\label{tab:ad_results}
\begin{tabular}{ccl|c|c|c|c|}
\cline{4-7}
\multicolumn{1}{l}{}                        & \multicolumn{1}{l}{}                        &           & \multicolumn{1}{c|}{\begin{tabular}[c]{@{}c@{}}Only \\ time-series\\ conversion \\ component\end{tabular}} & \multicolumn{1}{c|}{\begin{tabular}[c]{@{}c@{}}Using all \\ components\end{tabular}} & \multicolumn{1}{c|}{\begin{tabular}[c]{@{}c@{}}Using Feature\\ Engineering\\  component\end{tabular}} & \multicolumn{1}{c|}{\begin{tabular}[c]{@{}c@{}}Using Stationary\\ Conversion \\ component\end{tabular}} \\ \hline
\multicolumn{1}{|c|}{\multirow{6}{*}{DFL}}  & \multicolumn{1}{c|}{\multirow{3}{*}{fully}} & Precision & \textbf{0.9509}                                                                                            & 0.9075                                                                               & 0.9144                                                                                                & 0.9460                                                                                                  \\ 
\multicolumn{1}{|c|}{}                      & \multicolumn{1}{c|}{}                       & Recall    & 0.9369                                                                                                     & 0.8908                                                                               & 0.8982                                                                                                & 0.9326                                                                                                  \\ 
\multicolumn{1}{|c|}{}                      & \multicolumn{1}{c|}{}                       & F1-score  & 0.9405                                                                                                     & 0.8953                                                                               & 0.9068                                                                                                & 0.9366                                                                                                  \\ \cline{2-7} 
\multicolumn{1}{|c|}{}                      & \multicolumn{1}{c|}{\multirow{3}{*}{ring}}  & Precision & 0.9414                                                                                                     & 0.9108                                                                               & 0.9083                                                                                                & 0.9460                                                                                                  \\ 
\multicolumn{1}{|c|}{}                      & \multicolumn{1}{c|}{}                       & Recall    & 0.9266                                                                                                     & 0.8961                                                                               & 0.8948                                                                                                & 0.9324                                                                                                  \\ 
\multicolumn{1}{|c|}{}                      & \multicolumn{1}{c|}{}                       & F1-score  & 0.9306                                                                                                     & 0.9001                                                                               & 0.8985                                                                                                & 0.9364                                                                                                  \\ \hline
\multicolumn{1}{|c|}{\multirow{6}{*}{SDFL}} & \multicolumn{1}{c|}{\multirow{3}{*}{fully}} & Precision & 0.9507                                                                                                     & 0.9096                                                                               & 0.9389                                                                                                & 0.9389                                                                                                  \\ 
\multicolumn{1}{|c|}{}                      & \multicolumn{1}{c|}{}                       & Recall    & \textbf{0.9370}                                                                                            & 0.8971                                                                               & 0.9228                                                                                                & 0.9253                                                                                                  \\ 
\multicolumn{1}{|c|}{}                      & \multicolumn{1}{c|}{}                       & F1-score  & \textbf{0.9406}                                                                                            & 0.9006                                                                               & 0.9270                                                                                                & 0.9295                                                                                                  \\ \cline{2-7} 
\multicolumn{1}{|c|}{}                      & \multicolumn{1}{c|}{\multirow{3}{*}{ring}}  & Precision & 0.9452                                                                                                     & 0.9178                                                                               & 0.9188                                                                                                & 0.9396                                                                                                  \\ 
\multicolumn{1}{|c|}{}                      & \multicolumn{1}{c|}{}                       & Recall    & 0.9314                                                                                                     & 0.9062                                                                               & 0.9038                                                                                                & 0.9266                                                                                                  \\ 
\multicolumn{1}{|c|}{}                      & \multicolumn{1}{c|}{}                       & F1-score  & 0.9351                                                                                                     & 0.9094                                                                               & 0.9076                                                                                                & 0.9306                                                                                                  \\ \hline
\multicolumn{1}{|c|}{\multirow{3}{*}{CFL}}  & \multicolumn{1}{c|}{\multirow{3}{*}{star}}  & Precision & 0.9203                                                                                                     & 0.8929                                                                               & 0.8870                                                                                                & 0.9162                                                                                                  \\ 
\multicolumn{1}{|c|}{}                      & \multicolumn{1}{c|}{}                       & Recall    & 0.9033                                                                                                     & 0.8751                                                                               & 0.8720                                                                                                & 0.9009                                                                                                  \\ 
\multicolumn{1}{|c|}{}                      & \multicolumn{1}{c|}{}                       & F1-score  & 0.9082                                                                                                     & 0.8800                                                                               & 0.8762                                                                                                & 0.9057                                                                                                  \\ \hline
\end{tabular}
\end{table*}

\begin{figure}[tp]
	\centerline{\includegraphics[width=0.6\linewidth]{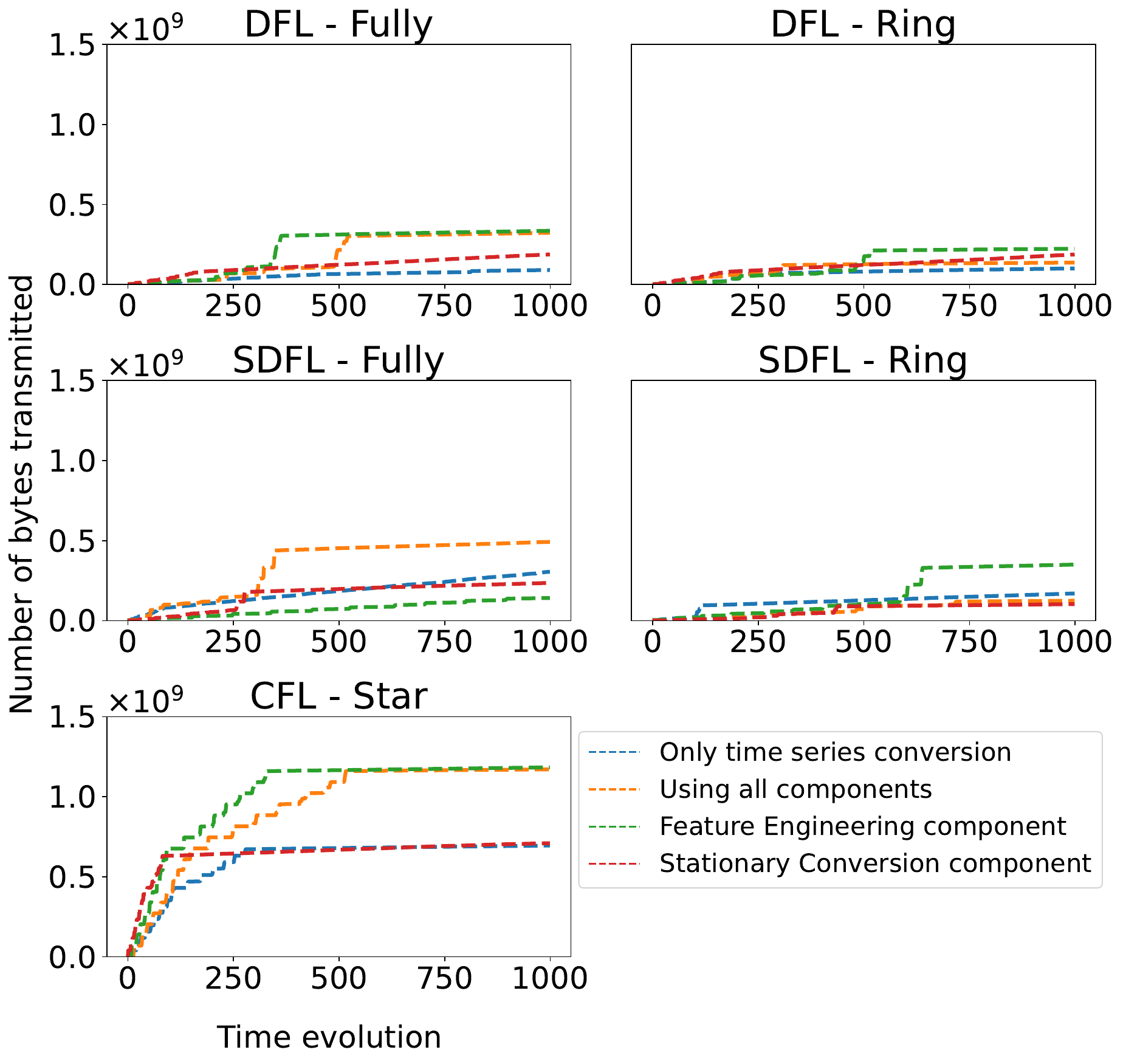}}
	\caption{Number of bytes transmitted for each TemporalFED configuration and under different paradigms and topologies}
	\label{fig:bytes_transmitted}
\end{figure}

\begin{figure}[tp]
	\centerline{\includegraphics[width=0.6\linewidth]{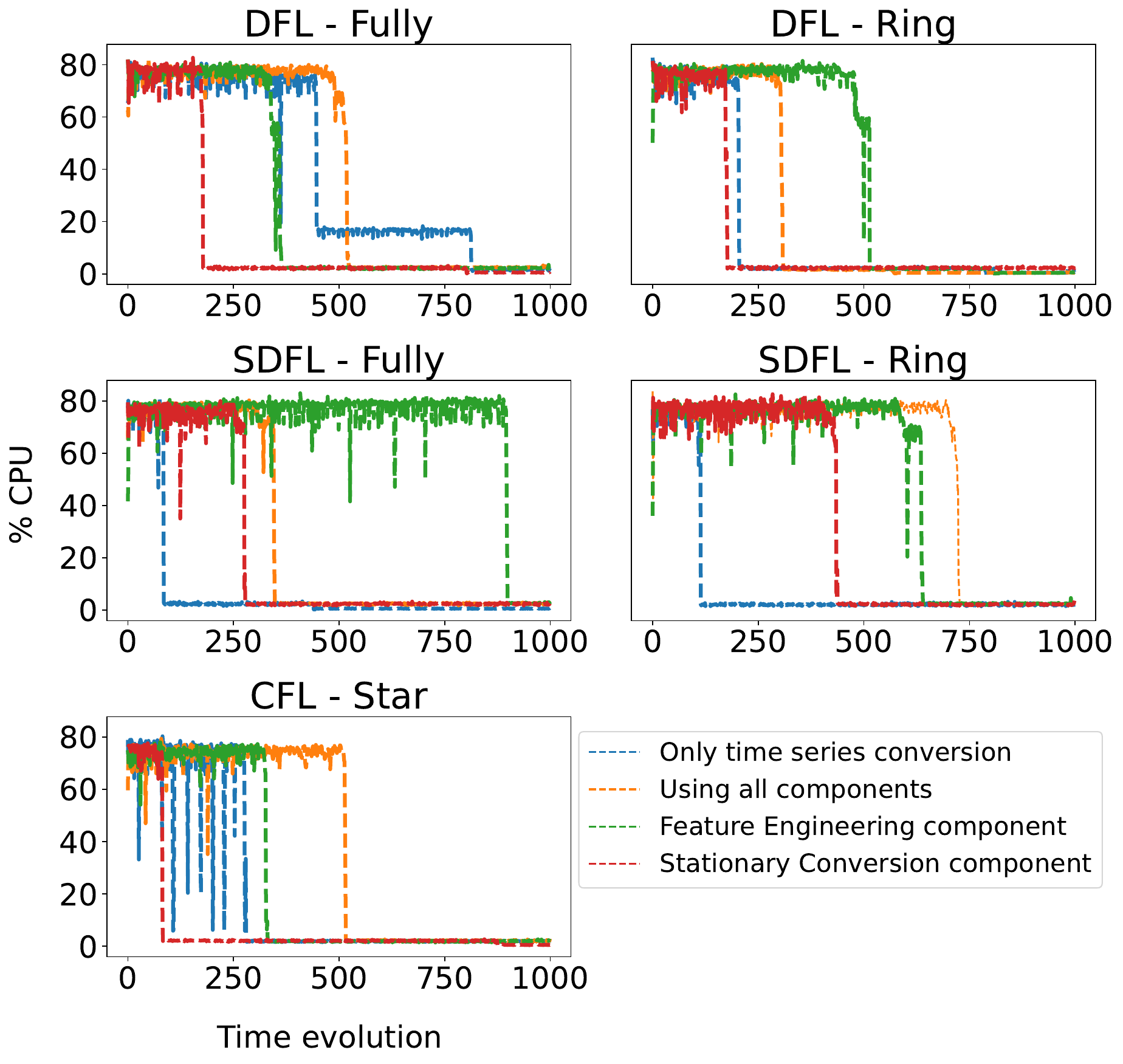}}
	\caption{Percentage of used CPU for each TemporalFED configuration and under different paradigms and topologies}
	\label{fig:cpu_usage}
\end{figure}

\begin{figure}[tp]
	\centerline{\includegraphics[width=0.6\linewidth]{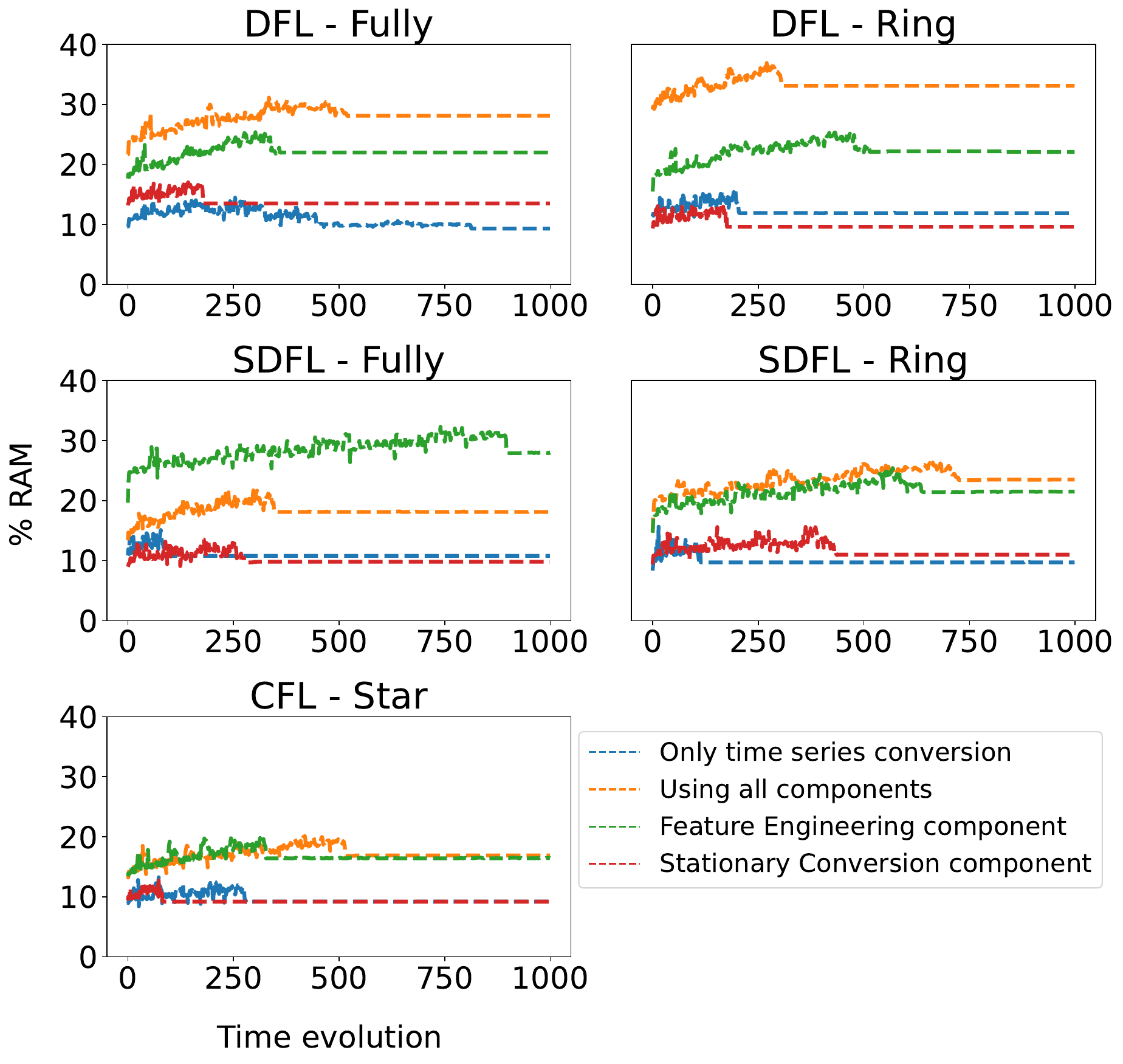}}
	\caption{Percentage of used RAM for each TemporalFED configuration and under different paradigms and topologies}
	\label{fig:ram_usage}
\end{figure}

In relation to resource consumption, it was examined the behavior across various paradigms, topologies, and configurations of the TemporalFED module in \figurename~\ref{fig:bytes_transmitted}, \figurename~\ref{fig:cpu_usage}, and \figurename~\ref{fig:ram_usage}. Each plot in these figures represents a paradigm and topology configuration, while the color of the lines within the plot represents the different configurations of the TemporalFED.

\figurename~\ref{fig:bytes_transmitted} presents the number of exchanged bytes, i.e., the sum of sent and received bytes, throughout the entire experiment. Clearly, the configuration of the TemporalFED utilizing feature extraction through DFT and AC (green and yellow lines) involves the most information exchange among participants since the number of features and weights being exchanged is higher than in other configurations. Notably, the number of exchanged bytes is particularly significant in the CFL paradigm compared to other scenarios. In particular, CFL using the TemporalFED configured with feature engineering component achieved $1.20 \times 10^{9}$ bytes transmitted, while in other configurations and paradigms and topology, the number of bytes exchanges varies from $0.25 \times 10^9$ to $0.50 \times 10^9$.

\figurename~\ref{fig:cpu_usage} illustrates the percentage of CPU usage during each experiment. In this case, all experiments utilize approximately 80\% of the CPU. It is worth noting that experiments utilizing the time-series conversion component (blue line) or the stationary conversion component (red line) required the least CPU usage, which aligns with their lower computational demands. Conversely, configurations employing the feature engineering component (yellow and green lines) utilized the CPU longer. Notably, the experiment using the SDFL paradigm, a fully connected topology, and solely the feature extraction component maintained nearly constant CPU usage throughout the entire experiment.

Finally, \figurename~\ref{fig:ram_usage} displays the percentage of RAM consumption during the execution of the experiments. This consumption pattern mirrors what was observed in the CPU usage. Specifically, configurations utilizing the feature extraction component (yellow and green lines) exhibited higher levels of RAM usage. In this context, the higher consumption was achieved by the DFL paradigm and ring topology, achieving a 30\% of RAM using all components and around 25\% of RAM using only the feature engineering component. In contrast, those configurations that did not employ this module (blue and red lines) made more efficient use of RAM. In this scenario, the RAM was between 10\% and 15\% for every experiment. Of special mention is the CFL paradigm, which required the least RAM consumption among configurations using the feature extraction component.

In conclusion, the resource consumption analysis sheds light on the varying demands imposed by different configurations of the TemporalFED module, providing valuable insights for optimizing performance and resource allocation in anomaly detection experiments.

\section{Conclusion and Future Work}
\label{sec:conclusions}

This work proposes TemporalFED, a software module for DFL platforms that analyzes time-series data to detect cyberattacks affecting industrial scenarios. TemporalFED comprises three components: the first converts tabular data into time-series data, suitable for Deep Learning algorithms; the second extracts features using DFT and autocorrelation techniques, while the third component converts time series into stationary form to facilitate processing. TemporalFED has been deployed on Fedstellar, a DFL platform enabling decentralized training of FL models under different configurations of clients, topologies, and algorithms. TemporalFED has been validated in a simulated yet realistic industrial testbed that gathers data from chemical industrial processes. The performance in terms of anomaly detection and resource consumption (bandwidth, \%CPU, and \%RAM) was evaluated in a pool of 20 experiments with different configurations in terms of FL paradigms (DFL, SDFL, and CFL), topologies (fully connected, ring, and star), and time series (solely time-series conversion, feature engineering alone, stationary conversion alone, and all three components combined). The results demonstrated that DFL and SDFL paradigms, in conjunction with the fully connected topology, achieved the best performance in anomaly detection. Particularly, the DFL combined with fully connected achieved 0.9509, 0.9369, and 0.9405 in precision, recall, and F1-score, respectively, while the SDFL and fully connected combination obtained 0.9507, 0.9370, and 0.9406 in precision, recall, and F1-score, respectively. Regarding resource consumption, the CFL paradigm exchanged the highest number of bytes with federation participants. CPU usage remained consistent at 80\% for all configurations, while RAM usage varied between 10\% and 35\% depending on the paradigm, topology, and configuration of the TemporalFED. In conclusion, the configuration using the feature engineering component consumed the highest amount of resources. 

As a future work direction, the robustness against adversarial attacks in distributed environments will be studied, such as DFL, incorporating explainability in the predictions made by distributed approaches.

\section{Acknowledgments}

This work has been funded under (a) Grant TED2021-129300B-I00, by MCIN/AEI/10.13039/501100011033, NextGeneration EU/PRTR, UE, (b) Grant PID2021-122466OB-I00, by MCIN/AEI/10.13039/501100011033/FEDER, UE, (c) the strategic project DEFENDER from the Spanish National Institute of Cybersecurity (INCIBE), by the Recovery, Transformation and Resilience Plan, Next Generation EU, (d) 21629/FPI/21, Fundación Séneca, Región de Murcia (Spain), (e) the Swiss Federal Office for Defense Procurement (armasuisse) with the DATRIS and CyberForce (CYD-C-2020003), and (f) by the University of Zurich (UZH).


\begin{thebibliography}{10}
\providecommand{\url}[1]{#1}
\csname url@samestyle\endcsname
\providecommand{\newblock}{\relax}
\providecommand{\bibinfo}[2]{#2}
\providecommand{\BIBentrySTDinterwordspacing}{\spaceskip=0pt\relax}
\providecommand{\BIBentryALTinterwordstretchfactor}{4}
\providecommand{\BIBentryALTinterwordspacing}{\spaceskip=\fontdimen2\font plus
\BIBentryALTinterwordstretchfactor\fontdimen3\font minus
  \fontdimen4\font\relax}
\providecommand{\BIBforeignlanguage}[2]{{%
\expandafter\ifx\csname l@#1\endcsname\relax
\typeout{** WARNING: IEEEtran.bst: No hyphenation pattern has been}%
\typeout{** loaded for the language `#1'. Using the pattern for}%
\typeout{** the default language instead.}%
\else
\language=\csname l@#1\endcsname
\fi
#2}}
\providecommand{\BIBdecl}{\relax}
\BIBdecl

\bibitem{ghobakhloo2020industry}
M.~Ghobakhloo, ``Industry 4.0, digitization, and opportunities for
  sustainability,'' \emph{Journal of cleaner production}, vol. 252, p. 119869,
  2020.

\bibitem{corallo2020cybersecurity}
A.~Corallo, M.~Lazoi, and M.~Lezzi, ``Cybersecurity in the context of industry
  4.0: A structured classification of critical assets and business impacts,''
  \emph{Computers in industry}, vol. 114, p. 103165, 2020.

\bibitem{Rey:2021:FL}
V.~Rey, P.~M. {Sánchez Sánchez}, A.~{Huertas Celdrán}, and G.~Bovet,
  ``Federated learning for malware detection in iot devices,'' \emph{Computer
  Networks}, vol. 204, p. 108693, 2022.

\bibitem{hao2019efficient}
M.~Hao, H.~Li, X.~Luo, G.~Xu, H.~Yang, and S.~Liu, ``Efficient and
  privacy-enhanced federated learning for industrial artificial intelligence,''
  \emph{IEEE Transactions on Industrial Informatics}, vol.~16, no.~10, pp.
  6532--6542, 2019.

\bibitem{MartinezBeltran:DFL_survey:2022}
E.~T. Mart{\'i}nez~Beltr{\'a}n, M.~Quiles~P{\'e}rez, P.~M.
  S{\'a}nchez~S{\'a}nchez, S.~L{\'o}pez~Bernal, G.~Bovet, M.~Gil~P{\'e}rez,
  G.~Mart{\'i}nez~P{\'e}rez, and A.~Huertas~Celdr{\'a}n, ``{Decentralized
  Federated Learning: Fundamentals, State-of-the-art, Frameworks, Trends, and
  Challenges},'' \emph{arXiv preprint arXiv:2211.08413}, 2022.

\bibitem{tahir2021experience}
B.~Tahir, A.~Jolfaei, and M.~Tariq, ``Experience-driven attack design and
  federated-learning-based intrusion detection in industry 4.0,'' \emph{IEEE
  Transactions on Industrial Informatics}, vol.~18, no.~9, pp. 6398--6405,
  2021.

\bibitem{MartinezBeltran:fedstellar:2023}
E.~T. Mart{\'i}nez~Beltr{\'a}n, A.~L. Perales~G\'omez, C.~Feng, P.~M.
  S{\'a}nchez~S{\'a}nchez, S.~L{\'o}pez~Bernal, G.~Bovet, M.~Gil~P{\'e}rez,
  G.~Mart{\'i}nez~P{\'e}rez, and A.~Huertas~Celdr{\'a}n, ``{Fedstellar: A
  Platform for Decentralized Federated Learning},'' \emph{arXiv preprint
  arXiv:2306.09750}, 2023.

\bibitem{Savazzi:dfl_industry_networks:2020}
S.~Savazzi, S.~Kianoush, V.~Rampa, and M.~Bennis, ``A joint decentralized
  federated learning and communications framework for industrial networks,'' in
  \emph{2020 IEEE 25th International Workshop on Computer Aided Modeling and
  Design of Communication Links and Networks (CAMAD)}, 2020, pp. 1--7.

\bibitem{Ma:iot_selection_quantized:2021}
T.~Ma, H.~Wang, and C.~Li, ``Quantized distributed federated learning for
  industrial internet of things,'' \emph{IEEE Internet of Things Journal},
  vol.~10, no.~4, pp. 3027--3036, 2023.

\bibitem{Qu:bl_cognitive_computing_industry:2021}
Y.~Qu, S.~R. Pokhrel, S.~Garg, L.~Gao, and Y.~Xiang, ``A blockchained federated
  learning framework for cognitive computing in industry 4.0 networks,''
  \emph{IEEE Transactions on Industrial Informatics}, vol.~17, no.~4, pp.
  2964--2973, 2021.

\bibitem{Singh:bl_5g_dfl:2023}
S.~K. Singh, L.~T. Yang, and J.~H. Park, ``Fusionfedblock: Fusion of blockchain
  and federated learning to preserve privacy in industry 5.0,''
  \emph{Information Fusion}, vol.~90, pp. 233--240, 2023.

\bibitem{Ranathunga:dfl_bl_industry:2023}
T.~Ranathunga, A.~McGibney, S.~Rea, and S.~Bharti, ``Blockchain-based
  decentralized model aggregation for cross-silo federated learning in industry
  4.0,'' \emph{IEEE Internet of Things Journal}, vol.~10, no.~5, pp.
  4449--4461, 2023.

\bibitem{Yuan:decefl_framework:2021}
Y.~Yuan, J.~Liu, D.~Jin, Z.~Yue, T.~Yang, R.~Chen, M.~Wang, C.~Sun, L.~Xu,
  F.~Hua, Y.~Guo, X.~Tang, X.~He, X.~Yi, D.~Li, G.~Wen, W.~Yu, H.-T. Zhang,
  T.~Chai, S.~Sui, and H.~Ding, ``Decefl: a principled fully decentralized
  federated learning framework,'' \emph{National Science Open}, vol.~2, no.~1,
  p. 20220043, 2023.

\bibitem{Qiu:dfl_industry:2023}
W.~Qiu, W.~Ai, H.~Chen, Q.~Feng, and G.~Tang, ``Decentralized federated
  learning for industrial iot with deep echo state networks,'' \emph{IEEE
  Transactions on Industrial Informatics}, vol.~19, no.~4, pp. 5849--5857,
  2023.

\bibitem{rodriguez2023anomaly}
M.~Rodr{\'\i}guez, D.~P. Tob{\'o}n, and D.~M{\'u}nera, ``Anomaly classification
  in industrial internet of things: A review,'' \emph{Intelligent Systems with
  Applications}, p. 200232, 2023.

\bibitem{nain2022towards}
G.~Nain, K.~Pattanaik, and G.~Sharma, ``Towards edge computing in intelligent
  manufacturing: Past, present and future,'' \emph{Journal of Manufacturing
  Systems}, vol.~62, pp. 588--611, 2022.

\bibitem{hiessl2020industrial}
T.~Hiessl, D.~Schall, J.~Kemnitz, and S.~Schulte, ``Industrial federated
  learning--requirements and system design,'' in \emph{International Conference
  on Practical Applications of Agents and Multi-Agent Systems}.\hskip 1em plus
  0.5em minus 0.4em\relax Springer, 2020, pp. 42--53.

\bibitem{liu2020deep}
Y.~Liu, S.~Garg, J.~Nie, Y.~Zhang, Z.~Xiong, J.~Kang, and M.~S. Hossain, ``Deep
  anomaly detection for time-series data in industrial iot: A
  communication-efficient on-device federated learning approach,'' \emph{IEEE
  Internet of Things Journal}, vol.~8, no.~8, pp. 6348--6358, 2020.

\bibitem{huong2021detecting}
T.~T. Huong, T.~P. Bac, D.~M. Long, T.~D. Luong, N.~M. Dan, B.~D. Thang, K.~P.
  Tran \emph{et~al.}, ``Detecting cyberattacks using anomaly detection in
  industrial control systems: A federated learning approach,'' \emph{Computers
  in Industry}, vol. 132, p. 103509, 2021.

\bibitem{truong2022light}
H.~T. Truong, B.~P. Ta, Q.~A. Le, D.~M. Nguyen, C.~T. Le, H.~X. Nguyen, H.~T.
  Do, H.~T. Nguyen, and K.~P. Tran, ``Light-weight federated learning-based
  anomaly detection for time-series data in industrial control systems,''
  \emph{Computers in Industry}, vol. 140, p. 103692, 2022.

\bibitem{du2023trans}
N.~H. Du, N.~H. Long, K.~N. Ha, N.~V. Hoang, T.~T. Huong, and K.~P. Tran,
  ``Trans-lighter: A light-weight federated learning-based architecture for
  remaining useful lifetime prediction,'' \emph{Computers in Industry}, vol.
  148, p. 103888, 2023.

\bibitem{jahromi2023ensemble}
A.~N. Jahromi, H.~Karimipour, and A.~Dehghantanha, ``An ensemble deep federated
  learning cyber-threat hunting model for industrial internet of things,''
  \emph{Computer Communications}, vol. 198, pp. 108--116, 2023.

\bibitem{de2023generalizing}
G.~de~Carvalho~Bertoli, L.~A.~P. Junior, O.~Saotome, and A.~L. dos Santos,
  ``Generalizing intrusion detection for heterogeneous networks: A
  stacked-unsupervised federated learning approach,'' \emph{Computers \&
  Security}, vol. 127, p. 103106, 2023.

\bibitem{masood2023blockchain}
A.~B. Masood, A.~Hasan, V.~Vassiliou, and M.~Lestas, ``A blockchain-based
  data-driven fault-tolerant control system for smart factories in industry
  4.0,'' \emph{Computer Communications}, vol. 204, pp. 158--171, 2023.

\bibitem{radaideh2022time}
M.~I. Radaideh, C.~Pappas, J.~Walden, D.~Lu, L.~Vidyaratne, T.~Britton,
  K.~Rajput, M.~Schram, and S.~Cousineau, ``Time series anomaly detection in
  power electronics signals with recurrent and convlstm autoencoders,''
  \emph{Digital Signal Processing}, vol. 130, p. 103704, 2022.

\bibitem{rieth2017issues}
C.~A. Rieth, B.~D. Amsel, R.~Tran, and M.~B. Cook, ``Issues and advances in
  anomaly detection evaluation for joint human-automated systems,'' in
  \emph{International Conference on Applied Human Factors and
  Ergonomics}.\hskip 1em plus 0.5em minus 0.4em\relax Springer, 2017, pp.
  52--63.

\bibitem{onel2019nonlinear}
M.~Onel, C.~A. Kieslich, and E.~N. Pistikopoulos, ``A nonlinear support vector
  machine-based feature selection approach for fault detection and diagnosis:
  Application to the tennessee eastman process,'' \emph{AIChE Journal},
  vol.~65, no.~3, pp. 992--1005, 2019.

\bibitem{McMahan:communication_efficient:2016}
H.~B. McMahan, E.~Moore, D.~Ramage, S.~Hampson, and B.~A.~y. Arcas,
  ``Communication-efficient learning of deep networks from decentralized
  data,'' \emph{arXiv preprint arXiv:1602.05629}, 2016.

\bibitem{zhang2022r}
Zhang, W., Yu, F., Wang, X., Zeng, X., Zhao, H., Tian, Y., ... \& Li, Z. R $^{2} $ fed: resilient reinforcement federated learning for industrial applications. IEEE Transactions on Industrial Informatics, 19(8), 8829-8840, 2022


\end{thebibliography}
\end{document}